\documentclass[aps,prl,floatfix,twocolumn,superscriptaddress]{revtex4-1}
\usepackage{float}
\usepackage{makeidx}
\usepackage{graphicx,amsmath}
\usepackage{colordvi}
\usepackage[dvipsnames]{xcolor}
\usepackage{ulem}
\usepackage{bbold}
\usepackage{bbm}
\allowdisplaybreaks
\usepackage{hyperref}
\hypersetup{backref,pdfpagemode=FullScreen,colorlinks=true,breaklinks,urlcolor=blue,linkcolor=blue,citecolor=blue}

\def\black{\color{black}}

\usepackage{mathrsfs}

\begin{document}

\title{Bosonic Topological Excitations from the Instability of a Quadratic Band Crossing}

\author{Guang-Quan Luo}
\affiliation{Department of Physics and Shenzhen Institute for Quantum Science and Engineering, Southern University of Science and Technology, Shenzhen 518055, China}
\affiliation{School of Physics, Huazhong University of Science and Technology, Wuhan 430074, China}

\author{Andreas Hemmerich}
\affiliation{Institut f\"ur Laser-Physik, Universit\"at Hamburg, Luruper Chaussee 149, 22761 Hamburg, Germany}

\author{Zhi-Fang Xu}
\email{xuzf@sustc.edu.cn}
\affiliation{Department of Physics and Shenzhen Institute for Quantum Science and Engineering, Southern University of Science and Technology, Shenzhen 518055, China}

\begin{abstract}
We investigate the interaction-driven instability of a quadratic band crossing arising for ultracold bosonic atoms loaded into a two-dimensional optical lattice. We consider the case when the degenerate point becomes a local minimum of both crossing energy bands such that it can support a stable Bose-Einstein condensate. Repulsive contact interaction among the condensed bosons induces a spontaneously time-reversal symmetry broken superfluid phase and a topological gap is opened in the excitation spectrum. We propose two concrete realizations of the desired quadratic band crossing in lattices with either fourfold or sixfold rotational symmetries via suitable tuning of the unit cell leading to reduced Brillouin zones and correspondingly folded bands. In either case, topologically protected edge excitations are found for a finite system. 
\end{abstract}

\maketitle

Ultracold atoms in optical lattices with sufficient tunability constitute an ideal platform for studying many-body physic and in particular topological states of quantum matter~\cite{Gross2017,Goldman2016,Bloch2012,Bloch2008,Lewenstein2007,Jaksch2005}. A wide variety of lattice potentials can be realized experimentally via interfering laser beams, ranging from triangular~\cite{Becker2010,Struck2011}, honeycomb~\cite{Soltan-Panahi2011}, checkerboard~\cite{Sebby-Strabley2006,Wirth2011}, Lieb~\cite{Taie2015} to Kagome~\cite{Jo2012} geometries leading to a rich collection of single-particle band structures.  

Experimental breakthroughs with regard to the implementation of synthetic magnetic fields~\cite{Lin2009,Aidelsburger2013,Miyake2013,Aidelsburger2015,Mancini2015,Stuhl2015} and spin-orbit coupling~\cite{Lin2011,Zhang2012,Wang2012,Cheuk2012,Huang2016,Wu2016, Luo2016,Wu2017}  for ultracold atoms using techniques as laser-assisted tunneling~\cite{Aidelsburger2013,Miyake2013,Aidelsburger2015,Mancini2015, Stuhl2015,Lin2011,Zhang2012,Wang2012,Cheuk2012,Huang2016,Wu2016}, lattice shaking~\cite{Struck2011,Parker2013,Jotzu2014}, and magnetic-field-gradient pulses~\cite{Xu2013,Anderson2013, Struck2014,Goldman2014,Jotzu2015,Luo2016,Wu2017} have provided us with the necessary building blocks for the formation of topological band insulators. According to the symmetry classification for noninteracting fermions~\cite{Altland1997,Ryu2010,Hasan2010}, schemes for realizing topological Chern insulators share the common feature to explicitly break time-reversal symmetry (TRS) on the single-particle level.

Spontaneous symmetry breaking induced by interaction provides another promising mechanism for achieving topological phases~\cite{Raghu2008,Sun2009,Sun2012,Uebelacker2011,Chern2012,Tsai2015,WuHQ2016,Zhu2016, Volovik2003,Kallin2012,Liu2014,Xu2017,Martin2008,Chern2012,Xu2015}. For a two-dimensional (2D) fermionic system, arbitrary repulsive interaction renders the quadratic band crossing unstable towards  a quantum anomalous Hall phase with a spontaneous TRS breaking~\cite{Sun2009,Sun2012}. Attractive interaction among fermions can also generate topological chiral superfluids if it possesses effective higher-partial-wave character~\cite{Volovik2003,Kallin2012} or even in the case of an $s$-wave interaction in spin-imbalanced systems~\cite{Liu2014,Xu2017}. A spontaneous quantum Hall effect emerges for fermions moving on an interaction-driven spin texture background in a triangular lattice~\cite{Martin2008,Xu2015}.

\begin{figure}[b]
\centering
\includegraphics[width=\linewidth]{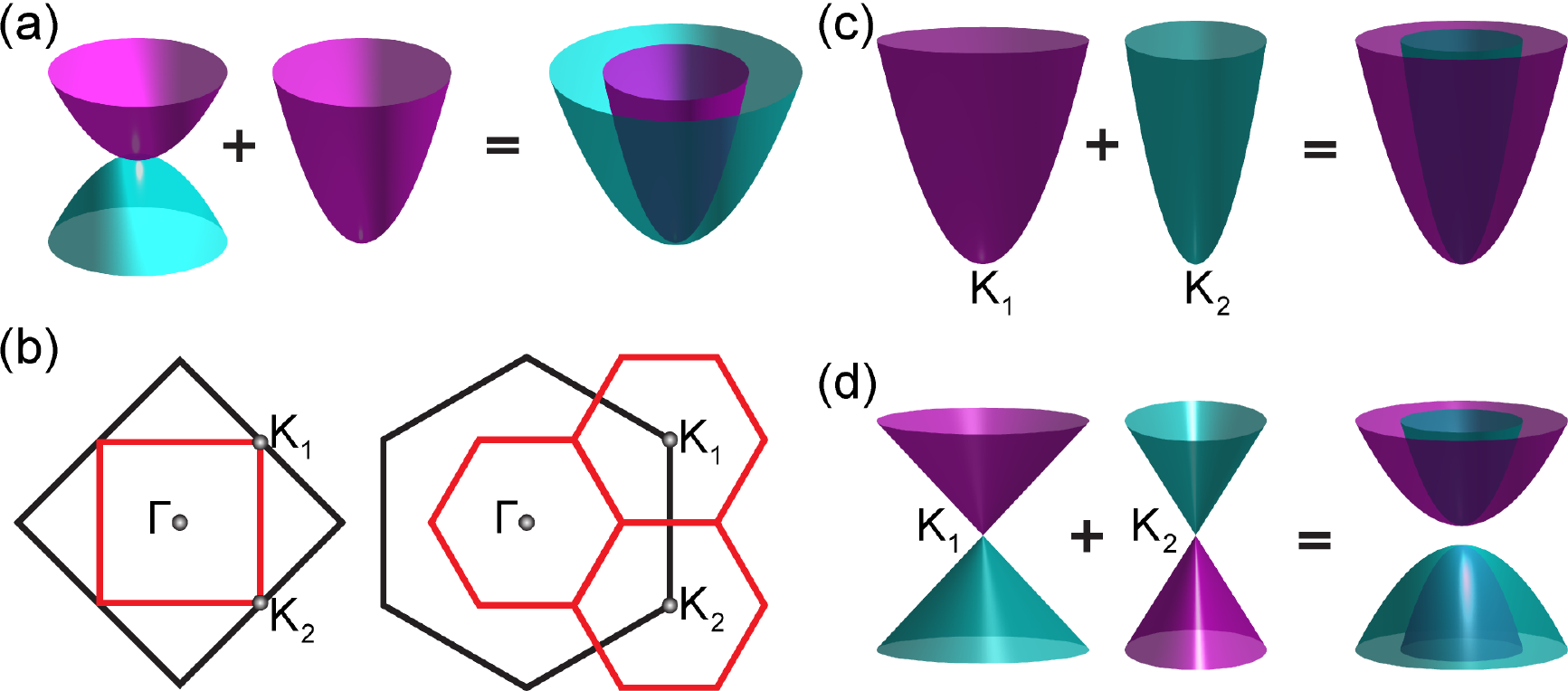}
\caption{ (color online). (a) Schematic picture for stabilizing bosons condensed at a quadratic band crossing via adding a parabolic single-particle spectrum. (b) Generic methods of realizing a quadratic band crossing by folding the Brillouin zone (BZ) in lattices with $D_4$ and $D_6$ point-group symmetries. The large black square (honeycomb) denotes the first BZ of the lattice, the small red square (honeycomb) shows the reduced BZ after folding. Momenta $K_1$ and $K_2$ are merged after folding. (c) Case of square lattice: Two parabolic band minima at $K_1$ and $K_2$ in the initial BZ are merged at the corner of the reduced BZ. (d) Case of honeycomb lattice: Two Dirac cones at $K_1$ and $K_2$ in the initial BZ are merged at the center ($\Gamma$-point) of the reduced BZ.}
\label{fig1}
\end{figure}

Recently, emergent topological phases in weakly interacting bosons have attracted great interest~\cite{Bardyn2016,Xu2016,Liberto2016,Wang2017}. Different methods of TRS breaking in a Bose-Einstein condensate (BEC) are utilized to implement a quantum Hall like effect for the bosonic excitations. Ref.~\cite{Bardyn2016} uses an optical phase imprinting method to create a background vortex lattice~\cite{Bardyn2016}. Refs.~\cite{Xu2016,Liberto2016} employ spontaneous TRS breaking induced by the Umklapp scattering among bosons in nearly degenerate $p$-orbitals in 2D optical lattices~\cite{Xu2016,Liberto2016}. However, each of these schemes  is only adapted to a specific lattice and the TRS breaking accompanied by a nonzero local or global orbital angular momentum of the condensate is insufficient to guarantee the emergence of topological excitations.  This has motivated us to identify the underlying fundamental mechanism, which allows us to give a more general recipe for realizing interaction-driven bosonic topological excitations.

In this work, we provide a general criterion for realizing topological excitations of Bose condensates~\cite{Bardyn2016,Xu2016,Liberto2016,Wang2017, Engelhardt2015,Furukawa2015,Pan2016}, that is 1. to provide a quadratic band crossing point (QBCP) and 2. a ferromagnetic orbital interaction to induce a spontaneous TRS breaking. In order to enable a stable BEC at the QBCP, it is chosen to be a local minimum of both crossing bands. Following this strategy, we point out two concrete experimentally realistic optical lattice scenarios with fourfold and sixfold rotational symmetries and single-particle spectra and interactions satisfying the requirements for having topological excitations. We thus extend previous studies on topological states in fermionic systems with a QBCP to bosonic systems. The fact that we only require $s$-wave interaction should facilitate experiments with bosonic systems as compared to their fermionic counterparts.

{\it Instability of the quadratic band crossing.}---
We consider a quasi two-dimensional (2D) optical lattice with a QBCP in single-particle energy spectra. In this work we focus on topological excitations of bosonic superfluids. In order to enable a stable BEC at the QBCP, the QBCP is assumed to form a local minimum of the crossing energy bands, as illustrated in Fig.~\ref{fig1}(a). We begin by formulating a low-energy theory for describing spinless bosons near a QBCP. We assume that the system is time-reversal invariant and loaded into lattices with a $D_4$ or $D_6$ point-group symmetry to protect the degeneracy of the QBCP~\cite{Sun2009}. Generally, we can use orbital bases of time-reversal pairs to formulate the theory, such as $p_x\pm ip_y$ and $d_{x^2-y^2}\pm id_{xy}$ orbitals. Close to the degenerate point, the low-energy Hamiltonian is given by~\cite{suppl}
\begin{eqnarray}
H&=&\int d\mathbf{r}\Big[{\bf \phi}^{\dagger}\mathcal{H}_0{\bf \phi}+\frac{g}{2}\sum\limits_{s=\pm}\phi_s^{\dagger}\phi_s^{\dagger} \phi_s\phi_s+2g\phi_+^{\dagger}\phi_-^{\dagger} \phi_-\phi_+\Big],\,\quad
\label{Hamiltonian}
\end{eqnarray}
where ${\bf \phi}^{\dagger}(\mathbf{r})=[\phi_+^{\dagger}(\mathbf{r}),\phi_-^{\dagger}(\mathbf{r})]$ and $\phi_{\pm}$ are field operators associated with time-reversal orbital pairs with $\mathcal{T}\phi_+\mathcal{T}^{-1}=\phi_-$ and $\mathcal{T}$ being the time-reversal operator.  A generic way to express the single-particle spectrum near a topologically non-trivial QBCP is~\cite{Sun2009}
\begin{eqnarray}
\mathcal{H}_0(\mathbf{k})=t_0(k_x^2+k_y^2)\sigma_0+t_1(k_x^2-k_y^2)\sigma_x+2t_2 k_xk_y\sigma_y,\quad
\label{qbcpham}
\end{eqnarray}
where $\sigma_0$ is the identity and $\sigma_{x,y}$ are Pauli matrices. Different from usual discussions on fermionic systems, we choose $t_0>\max(|t_1|,|t_2|)$ to ensure the stability of the BEC assumed to reside at the QBCP. Here, we consider an experimentally realizable ferromagnetic orbital interaction with $g>0$~\cite{Parker2013} as shown in Eq.~(\ref{Hamiltonian}), which naturally arises from a merely repulsive $s$-wave contact interaction. Its detailed form in Eq.~(\ref{Hamiltonian}) is obtained from a two-mode approximation and the requirement of orbital angular momentum conservation~\cite{suppl}. This interaction favors TRS breaking, when bosons condense at the QBCP~\cite{suppl}, leading to a quantum Hall like state as we clarify below.

We choose a degenerate ground state with $\langle \phi^{\dagger}(\mathbf{k}=\mathbf{0})\rangle=[\sqrt{N}, 0]$, where all $N$ bosons condense in the $\phi_+$ mode at zero momentum, and apply the number-conserving approach to obtain excitations~\cite{suppl}. The Hamiltonian is written as
$H=\frac{1}{2}\sum_{\mathbf{k}}({\bf \phi}_{\mathbf{k}}^{\dagger}, {\bf \phi}_{-\mathbf{k}})\mathcal{H}_{\rm BdG}(\mathbf{k}) ({\bf \phi}_{\mathbf{k}}, {\bf \phi}_{-\mathbf{k}}^{\dagger})^T$. The Bogoliubov-de Gennes (BdG) Hamiltonian reads
\begin{eqnarray}
\mathcal{H}_{\rm BdG}(\mathbf{k})=\sigma_0\otimes[\mathcal{H}_0(\mathbf{k})+gn\sigma_0]+\sigma_x\otimes [\sigma_z+\sigma_0]\frac{gn}{2}.\quad
\end{eqnarray}
where $n=N/V$ and $V$ are the density of bosons and the volume of the 2D system, respectively.

To satisfy the bosonic commutation relation, the BdG Hamiltonian should be diagonalized by a paraunitary matrix as $T^{\dagger}_{\mathbf{k}}\mathcal{H}_{\rm BdG}(\mathbf{k})T_{\mathbf{k}}=E_{\mathbf{k}}$, where $T^{\dagger}_{\mathbf{k}}\tau_zT_{\mathbf{k}}=\tau_z$ and $\tau_z=\sigma_z\otimes\sigma_0$. The topological behavior of excitations for the $j$-th band is then characterized by $\mathcal{C}_j=(1/2\pi)\int d\mathbf{k} B_j(\mathbf{k})$~\cite{Avron1983,Shindou2013}, where $B_j(\mathbf{k})\equiv\partial_{k_x}A_{j,y}(\mathbf{k})-\partial_{k_y}A_{j,x}(\mathbf{k})$, $A_{j,\nu}(\mathbf{k})\equiv i{\rm Tr}[\Gamma_j \tau_zT^{\dagger}_{\mathbf{k}}\tau_z\partial_{k_{\nu}}T_{\mathbf{k}}]$ and $\Gamma_j$ is a diagonal matrix with the $j$-th diagonal term equal to 1 and other terms are 0. 

\begin{figure}[tbp]
\centering
\includegraphics[width=0.9\linewidth]{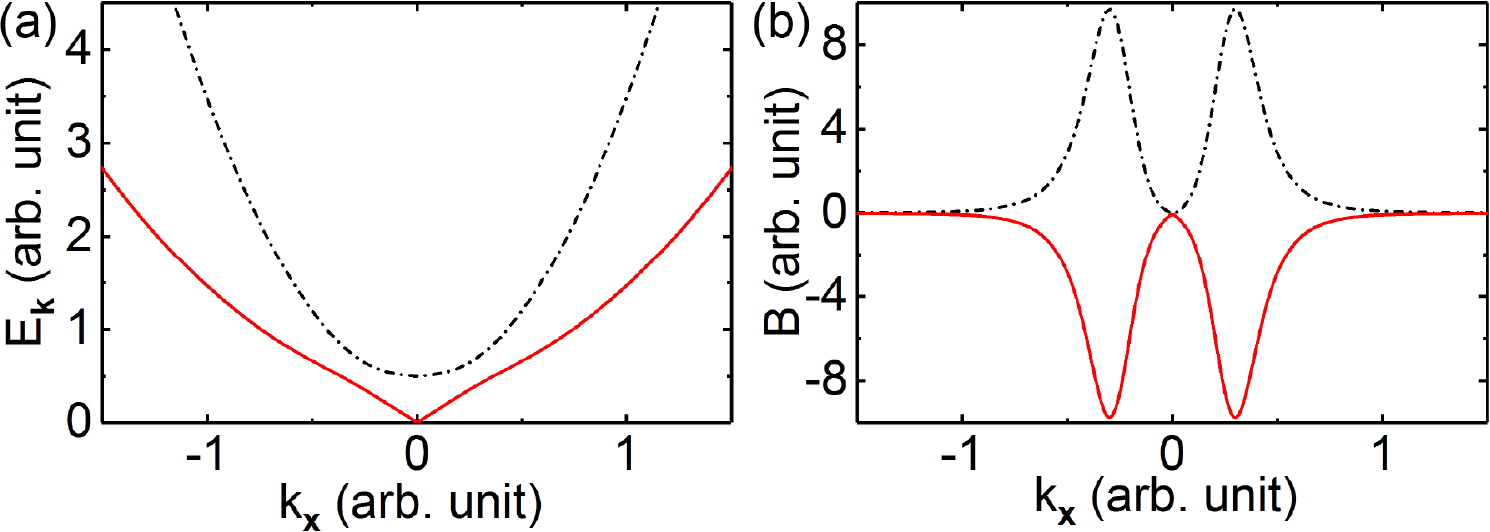}
\caption{ (color online). (a) Excitations for the TRS broken ground state with $\langle \phi^{\dagger}(\mathbf{k}=\mathbf{0})\rangle=[\sqrt{N},0]$. (b) Berry curvatures for the lowest (red solid line) and the first (black dashed-dotted line) band of the Bogoliubov excitation spectrum. Here, we choose $t_0=2t_1=2t_2=2$ and $gn=0.5$.}
\label{fig2}
\end{figure}

We find that as long as the TRS is broken by the ferromagnetic orbital interaction, a topological gap is opened at the QBCP. Figure~\ref{fig2} shows the first two excitation bands and the corresponding Berry curvatures. Numerically, we find that $\mathcal{C}_1=-1$ and $\mathcal{C}_2=1$ for the lowest and the first excited bands. Due to the bulk-boundary correspondence, there are topological edge modes within the gap for the finite system to be described below. We have thus pointed out a general scenario for realizing topological excitations in a bosonic system with a QBCP. In the following, we further propose two generic methods to realize the desired QBCP in optical lattices by folding the Brillouin zone in lattices with $D_4$ or $D_6$ point-group symmetries, as illustrated in Fig.~\ref{fig1}. One relies on superposing two parabolic band dispersions and the other involves the coupling between two Dirac cones.

{\it $D_4$ symmetry.}---
We first consider a 2D optical lattice preserving the $D_4$ point-group symmetry. The system thus is invariant under the parity operation $\mathcal{P}$. As illustrated in Fig.~\ref{fig1}(b), the single-particle states at inversion-invariant points $K_1$ and $K_2$ have defined parities. We further consider the energy band where single-particle states at $K_1$ and $K_2$ have odd parities and the energy spectra close to these two points show parabolic dispersions. Under these conditions, we propose a general scheme to realize a QBCP by simply folding the Brillouin zone, as demonstrated in Fig.~\ref{fig1}(c). The key reason is that odd-parity states at $K_1$ and $K_2$ form two orthogonal bases for the 2D irreducible representation of the $D_4$ symmetry group, which is also the little group for two inversion-invariant points after folding. To protect the QBCP, we always guarantee that the Hamiltonian preserves the $D_4$ point-group symmetry. In the following, we consider a concrete example to demonstrate the idea.

\begin{figure}[tbp]
\centering
\includegraphics[width=\linewidth]{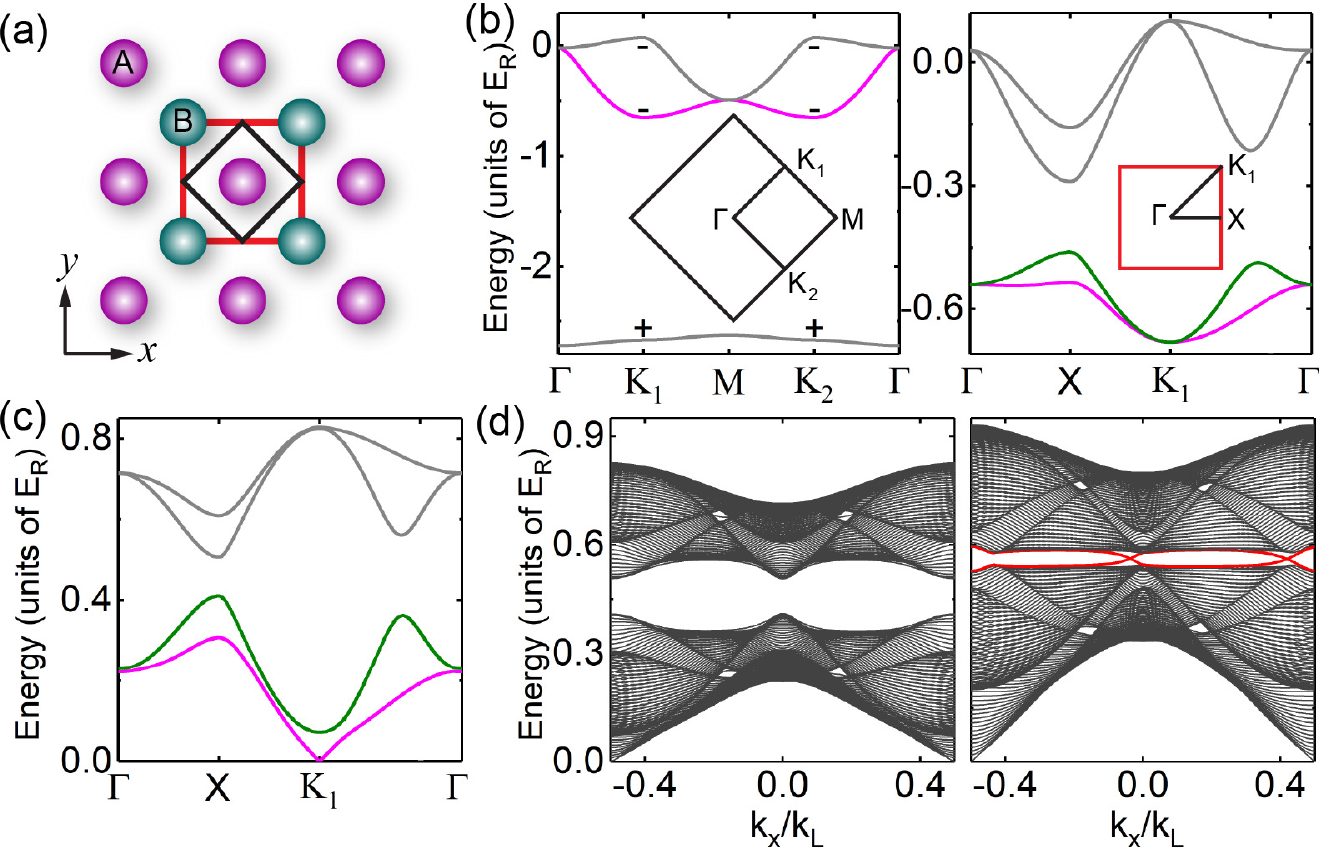}
\caption{ (color online). (a) Schematic of the square lattice geometry. The inner black solid and the outer red solid squares denote the unit cells of the lattice for $V_2 = 0$ and $V_2 \neq 0$, respectively. (b) Single-particle energy bands for $V_1=1.4\, E_R$. The left and right panels show the energy bands in the original first BZ with $V_2 = 0$ and in the reduced first BZ for $V_2 = 0.1 \, V_1$, respectively. The symbols `+' and `-' are used to denote even and odd parity of single-particle states at inversion-invariant points. (c) Bogoliubov excitation spectra for bosons condensed at the quadratic band crossing point. (d) Excitation spectra for a finite system with a cylinder geometry at a weak (left) and a larger (right) interaction. A periodic (open) boundary condition is assumed in the $x$ ($y$) direction. Solid black and red lines denote the bulk excitation spectra and chiral edge states, respectively.}
\label{fig3}
\end{figure}

We consider the lattice potential 
\begin{eqnarray}
V_{s} &=& -V_1\, [\cos k_L x+\cos k_L y]^2  \nonumber\\
 &+& 2 V_2 \, [\cos^2 (k_L x/2)  + \cos^2 (k_L y/2)]-2V_2. 
\end{eqnarray} 
The $V_1$ term is a special case of the chequerboard lattice realized in Ref.~\cite{Wirth2011} obtained by interfering two optical standing waves with wave numbers $k_L$. The $V_2$ term describes a conventional square lattice, obtained from the superposition of two non-interfering standing waves with wave numbers $k_L/2$. When $V_2=0$, the unit cell contains only a single lattice site hosting $s$, $p_x$ and $p_y$ orbitals with different parity. For $V_1=1.4\,E_R$ ($E_R\equiv\hbar^2k_L^2/2m$), the lowest three energy bands along highly symmetric lines in the first BZ are shown in Fig.~\ref{fig3}(b), where the minima of the second band related to $p$-orbitals are located at the $K_1$ and $K_2$ points. The corresponding single-particle states have odd parities. The second band shows a parabolic dispersion close to the two minima. Via turning on the $V_2$ term, the unit cell of the lattice is enlarged to comprise two lattice sites, and hence a reduced first BZ arises and each energy band of the $V_1$ lattice splits into two bands of the total lattice $V_{s}$. As long as the extra potential $V_2$ does not break the $D_4$ point-group symmetry, the two-fold degeneracy at $K_1$ ($K_2$) is not lifted after folding. This is illustrated in the right panel of Fig.~\ref{fig3}(b), where the four energy bands resulting from odd-parity $p$-orbitals are shown. We thus find a QBCP to appear at the corner of the reduced BZ, which is a local minimum of both bands involved and can support a stable BEC. 

To make our argument more quantitative, we consider a tight-binding model, which involves $p_x$ and $p_y$ orbitals at each lattice site. The single-particle Hamiltonian is 
\begin{eqnarray}
H_0&=&J_1\sum_{\mathbf{r},\mu}\left[\hat{p}_{\mu,\mathbf{r}}^\dag\hat{p}_{\mu,\mathbf{r} +\mathbf{e}_{\mu}+\mathbf{e}_{\bar{\mu}}} +\hat{p}_{\mu,\mathbf{r}}^\dag\hat{p}_{\mu,\mathbf{r}+\mathbf{e}_{\mu}-\mathbf{e}_{\bar{\mu}}}+\text{H.c.}\right]\nonumber\\
&+&J_2\sum_{\mathbf{r},\mu}\left[\hat{p}_{\mu,\mathbf{r}}^\dag\hat{p}_{\bar{\mu},\mathbf{r}+\mathbf{e}_{\mu}+\mathbf{e}_{\bar{\mu}}} -\hat{p}_{\mu,\mathbf{r}}^\dag\hat{p}_{\bar{\mu},\mathbf{r}+\mathbf{e}_{\mu}-\mathbf{e}_{\bar{\mu}}}+\text{H.c.}\right]\nonumber\\
&+&\sum_{\mathbf{r},\mu}\left[J_3\,\hat{p}_{\mu,\mathbf{r}}^\dag\hat{p}_{\mu,\mathbf{r}+2\mathbf{e}_\mu}
+J_4\,\hat{p}_{\mu,\mathbf{r}}^\dag\hat{p}_{\mu,\mathbf{r}+2\mathbf{e}_{\bar{\mu}}}
+\text{H.c.})\right]\nonumber\\
&+&\delta\sum_{\mathbf{r}\in A,\mu}\hat{p}_{\mu,\mathbf{r}}^\dag\hat{p}_{\mu,\mathbf{r}} -\delta\sum_{\mathbf{r}\in B,\mu}\hat{p}_{\mu,\mathbf{r}}^\dag\hat{p}_{\mu,\mathbf{r}}.
\end{eqnarray}
Here, $\hat{p}_{\mu,\mathbf{r}_{\alpha}}$ denotes the bosonic annihilation operator for the $p_{\mu}$ orbital located at $\mathbf{r}$, where $\mu=\{x,y\}$, $\bar{\mu}=\{\bar{x},\bar{y}\}=\{y,x\}$. $\mathbf{e}_{x}=(a/2,0)$, $\mathbf{e}_y=(0,a/2)$, and $a=2\pi/k_L$ is the wavelength used to form the $V_1$-lattice. The $\delta$-term arises for $V_2 \ne 0$, which yields a decomposition of the lattice into two sub-lattices denoted $A$ and $B$ with an on-site energy difference $\delta$ for $p$-orbitals at $A$ and $B$ sites. 

To obtain an effective Hamiltonian close to the QBCP, we project the Hamiltonian into a subspace spanned by two bases $\hat{P}_{\pm,\mathbf{k}_{K_1}}^\dag=(\pm i\hat{p}_{A\mp,\mathbf{k}}^\dag\cos\chi +\hat{p}_{B\pm,\mathbf{k}}^\dag\sin\chi)/\sqrt{2}$, where subscripts A and B are used to distinguish annihilation operators for orbital located at two different lattice sites, $\mathbf{k}_{K_1}=\mathbf{k}-(\pi/a,\pi/a)$, $\hat{p}_{\pm}^\dag=\hat{p}_x^\dag\pm i\hat{p}_y^\dag$, and $\chi=\arctan[(\delta+\sqrt{\delta^2+16J_2^2})/4J_2]$. The resulting effective single-particle Hamiltonian is given by $\tilde{H}_0=\sum_{\mathbf{k}}(\hat{P}_{+,\mathbf{k}},\hat{P}_{-,\mathbf{k}}^{\dagger}) \mathcal{H}_0(\mathbf{k})(\hat{P}_{+,\mathbf{k}},\hat{P}_{-,\mathbf{k}})^T$, where $\mathcal{H}_0(\mathbf{k})$ is given by Eq.~(\ref{qbcpham}) with parameters $t_0=a^2(J_3+J_4+J_2\sin2\chi)/2$, $t_1=a^2(J_4-J_3)\cos2\chi/2$, and $t_2=a^2J_1\sin2\chi$. The effective interaction part of the Hamiltonian takes the same form as that shown in Eq.~(\ref{Hamiltonian})~\cite{suppl}. This guarantees that we obtain a TRS broken phase.

{\it $D_6$ symmetry.}---
For the optical lattice with a $D_6$ point-group symmetry, we propose another mechanism to realize a QBCP shown in Fig.~\ref{fig1}(d). The basic idea is to interfere two Dirac cones at $K_1$ and $K_2$ via folding the Brillouin zone. Since the system preserves the $D_6$ symmetry, there is a symmetry-protected two-fold degeneracy. To be more specific, we consider an effective Hamiltonian for describing the energy dispersion close to two Dirac cones as
\begin{eqnarray}
\mathcal{H}_0(\mathbf{k})=v(k_y\nu_z\sigma_x-k_x\nu_0\sigma_y),
\label{diracham}
\end{eqnarray}
where $v$ is the velocity for two Dirac cones and $\nu_i$ are Pauli matrices used to describe the valley degrees of freedom. To open a gap at the Dirac cones, we can seek for operators that anticommute with $\mathcal{H}_0(\mathbf{k})$. Four operators $\nu_0\sigma_z$, $\nu_z\sigma_z$, $\nu_x\sigma_x$ and $\nu_y\sigma_z$ are found to be anticommuting with both $\nu_z\sigma_x$ and $\nu_0\sigma_y$~\cite{Liu2017}. Here, we focus on the perturbation $\delta\mathcal{H}_0=\Delta \tau_x\sigma_x$, which is found to be the key for generating $Z_2$ topological states associated with a pseudo TRS originating from the orbital rotation~\cite{WuLH2015,WuLH2016}. In contrast, we are mainly interested in the band dispersion, which contains two doubly degenerate bands with energy spectra $\varepsilon(\mathbf{k})=\pm\sqrt{v^2\mathbf{k}^2+\Delta^2}$. Taking into account quadratic terms ($\propto k_x^2, k_y^2, k_xk_y$) during deriving the effective Hamiltonian of Eq.~(\ref{diracham}), the full-band degeneracy should be lifted leaving only a QBCP at the $\Gamma$ point, as is illustrated in Fig.~\ref{fig1}(d).

A suitable optical lattice potential can be generated by overlapping two honeycomb lattices indexed by $\xi=s,l$. Each of these lattices is created by three laser beams intersecting pairwise under angles of $2\pi/3$ in the $xy$-plane. Their wave vectors are $\mathbf{k}_{\xi j}=k_{\xi} \, (\cos\theta_{\xi j},\sin\theta_{\xi j})$ where $j=1,2,3$, $k_{s} = \sqrt{3} \, k_{l}=k_L$, $\theta_{sj}=\pi/2+2\pi j/3$, and $\theta_{lj}=2\pi j/3$. These laser beams interfere to produce the optical lattice potential 
\begin{eqnarray}
V_{2h}=V_{s}\sum_j\cos\mathbf{b}_{sj}\cdot\mathbf{r}+ V_{l} \sum_j\cos\mathbf{b}_{lj}\cdot\mathbf{r},
\end{eqnarray}
where $\mathbf{b}_{\xi 1}=\mathbf{k}_{\xi  2}-\mathbf{k}_{\xi 3}$,
$\mathbf{b}_{\xi 2}=\mathbf{k}_{\xi 3}-\mathbf{k}_{\xi 1}$,
$\mathbf{b}_{\xi 3}=\mathbf{k}_{\xi 1}-\mathbf{k}_{\xi 2}$.

\begin{figure}[tbp]
\centering
\includegraphics[width=\linewidth]{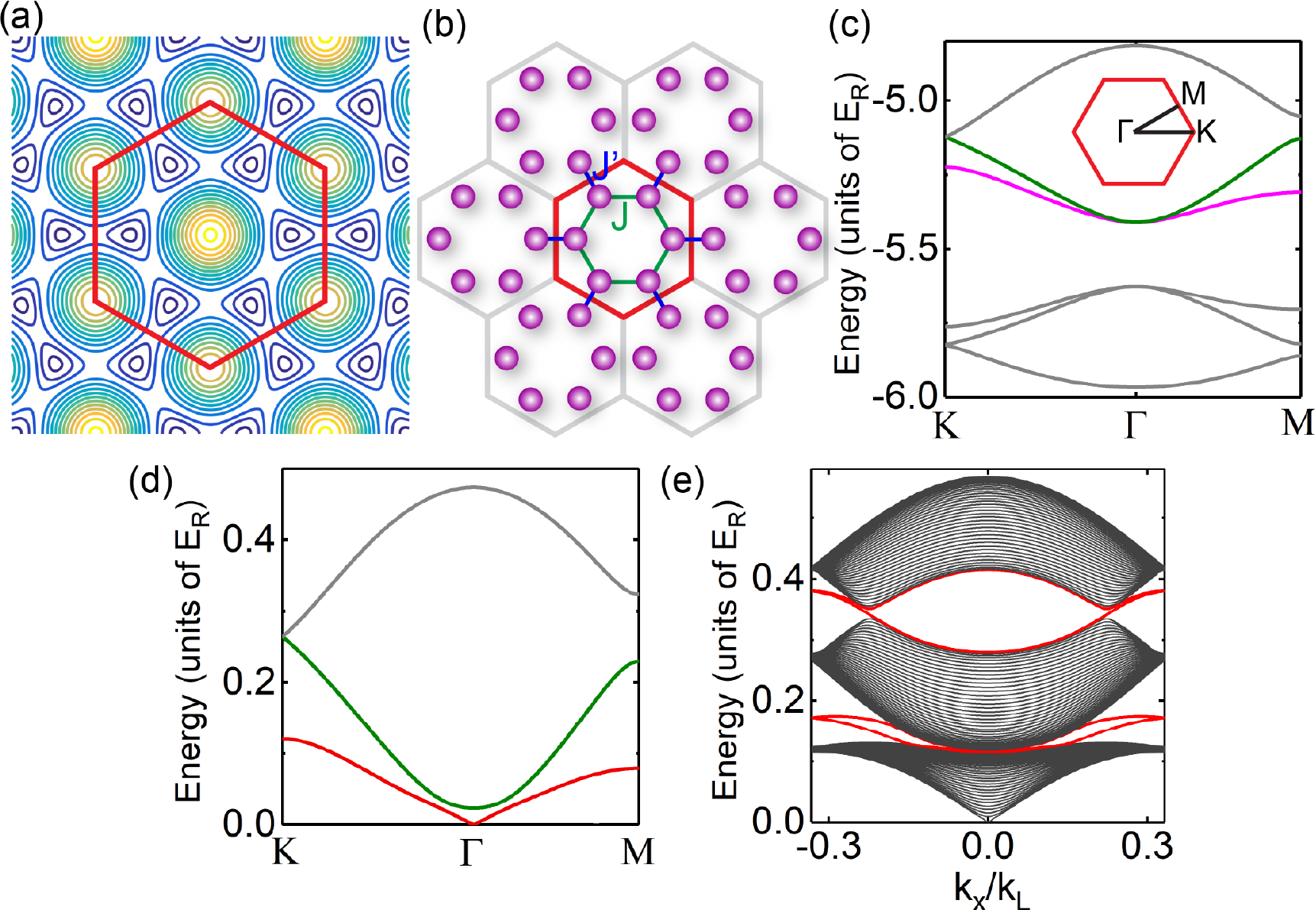}
\caption{ (color online). (a) Contour plot of the lattice potential $V_{2h}$. Red solid honeycomb denotes one unit cell of the lattice. (b) Schematic picture for the lattice structure. Larger honeycomb represents one unit cell. $J$ and $J'$ denote nearest-neighbor hoppings among two sites in the same and different unit cells, respectively. (c) The energy band dispersions along highly symmetric lines for the case with $V_{s} = 7\, E_R$ and $V_{l} = 0.1\, V_{s}$. (c) Bogoliubov excitation spectra for bosons condensed at the quadratic band crossing point. (d) Excitation spectra for a finite system with a cylinder geometry. Solid red lines denote the chiral edge states.}
\label{fig4}
\end{figure}

The $V_{s}$-term denotes the honeycomb lattice previously realized experimentally in Ref.~\cite{Soltan-Panahi2011}. Its unit cell covers two lattice sites. The addition of the larger honeycomb lattice described by the $V_2$ term increases the unit cell to comprise six lattice sites. As a consequence, the two Dirac cones at $K_1$ and $K_2$ in the first BZ of the $V_{s}$-lattice merge at the $\Gamma$ point of the reduced first BZ of the total lattice $V_{2h}$. \black The coupling among two valleys leads to a bulk gap opening, as illustrated in Fig.~\ref{fig4}(c). Since both honeycomb lattices preserve the $D_6$ point-group symmetry, the two-fold degeneracy at the $\Gamma$ point is protected. We then apply a tight-binding model to clarify the details. When $V_{l}/V_{s} \ll 1$, we only need to consider the nearest-neighbour hopping amplitudes. The corresponding single-particle Hamiltonian is
\begin{eqnarray}
H_0=\sum_{\langle i,j\rangle} \left(J_{ij}\hat{s}_{i}^\dag\hat{s}_{j}+\text{H.c.}\right).
\end{eqnarray}
Here, $\hat{s}_{i}$ denote the bosonic annihilation operator for the $s$ orbital at site $i$ and $\langle i,j\rangle$ denotes two nearest neighbor sites $i$ and $j$. $J_{ij}=J$ and $J_{ij}=J'\equiv\gamma J$ are the nearest-neighbor hopping coefficients between two sites in the same and in different unit cells, respectively, as illustrated in Fig.~\ref{fig4}(b). Two QBCPs are created at the $\Gamma$ point when $\gamma\neq 1$. We focus on the upper one where the degenerate point is a local minimum.

Using the projection operators of the $D_6$ group, we construct six symmetry-distinguished bases denoted as $s$, $p_x$, $p_y$, $d_{x^2-y^2}$, $d_{xy}$ and $f$ from six $s$-orbitals in one unit cell.
Close to the $\Gamma$ point, the middle four of the six resulting bands are spanned by $p_x$, $p_y$, $d_{x^2-y^2}$ and $d_{xy}$-orbitals. The corresponding single-particle Hamiltonian is $H_0=\sum_\mathbf{k}\Psi_\mathbf{k}^\dag \mathcal{H}_0(\mathbf{k})\Psi_\mathbf{k}$. Here, $\Psi_\mathbf{k}=(\hat{d}_{+,\mathbf{k}},\ \hat{d}_{-,\mathbf{k}},\ \hat{p}_{+,\mathbf{k}},\ \hat{p}_{-,\mathbf{k}})^\mathrm{T}$, $\hat{p}_\pm^\dag=\frac{1}{\sqrt{2}}( \hat{p}_x^\dag\pm i\hat{p}_y^\dag)$, $\hat{d}_\pm^\dag=\frac{1}{\sqrt{2}}( \hat{d}_{x^2-y^2}^\dag\pm i\hat{d}_{xy}^\dag)$. $\hat{p}_x$, $\hat{p_y}$, $\hat{d}_{x^2-y^2}$, and $\hat{d}_{xy}$ are annihilation operators for $p_x$, $p_y$, $d_{x^2-y^2}$, and $d_{xy}$-orbitals.

When $\gamma>1$, the upper QBCP arises from two $p$-orbitals. Projecting out two $d$-orbitals via a second-order perturbation, we obtain an effective Hamiltonian $\tilde{H}_0=\sum_{\mathbf{k}}(\hat{p}_{+,\mathbf{k}},\hat{p}_{-,\mathbf{k}}^{\dag}) \tilde{\mathcal{H}}_0(\mathbf{k})(\hat{p}_{+,\mathbf{k}},\hat{p}_{-,\mathbf{k}})^T$. We confirm that $\tilde{H}_0(\mathbf{k})$ has the same form of Eq.~(\ref{qbcpham}) with parameters $t_0=a_0^2(\gamma^2-8\gamma-2)J/8(\gamma-1)$ and $t_1=t_2=a_0^2(\gamma+2)J/8$, where $a_0$ is the distance between two nearest-neighbor sites. We also confirm that the interaction-part Hamiltonian is the same as that shown in Eq.~(\ref{Hamiltonian}). When $\gamma<1$, the upper QBCP originates from two $d$-orbitals. The energy dispersion has a similar form as in the case of $\gamma>1$.

{\it Topological excitations.}---
We further investigate bosonic excitations on top of the condensate beyond the low-energy effective model of Eq.~(\ref{Hamiltonian})~\cite{suppl}. We derive the hopping parameters via directly mapping out Wannier functions. It is confirmed that for both lattice potentials considered here a gap is opened at the degenerate point as long as TRS is broken due to the specific ferromagnetic orbital interaction, as illustrated in Fig.~\ref{fig3}(c) and Fig.~\ref{fig4}(d). This generates nonzero Berry curvatures~\cite{suppl}, which should lead to a Hall response of the system at a finite temperature. Recently developed techniques for measuring optical conductivity in cold atoms~\cite{Anderson2017} should be applicable for the experimental confirmation of the Hall effect. For the $D_6$ symmetric lattice, the low-energy effective model well describes the system and all predictions of the general treatment are reproduced. More specifically, topological protected edge modes emerge among the lowest two bands for a finite system. Furthermore, the TRS broken condensate also induces a topological bulk gap close to the Dirac cone at $K$ point, leading to in-gap edge modes for a finite system. For the $D_4$ symmetric square lattice, the single-particle spectra are degenerate not only at the QBCP we discussed ($K_1$ point) but also at the $\Gamma$ point. The Berry curvatures close to two points take different signs resulting in a zero topological invariant and no edge mode within the gap. However, when we increase the contact interaction, an interaction induced band inversion appears between the second and the third excitation band, which leads to a topologically nontrivial bulk gap and in-gap edge modes for a finite system, as illustrated in Fig.~\ref{fig3}(d)~\cite{suppl}.

{\it Conclusion.}---We proposed a generic method for realizing bosnic topological excitations based on the instability of a quadratic band crossing due to ferromagnetic orbital interaction among condensed bosons. The necessary QBCP can be realized via folding odd-parity orbital bands of a square lattice or interfering two Dirac cones in a honeycomb lattice. Moreover, we have unveiled the edge modes for the interaction-induced TRS broken phase. It is an interesting question whether the remarkable features we have found for weakly interacting bosonic superfluids can extend to the strongly interacting regime. Finally, the considerations of this work can also be applied for realizing similar bosonic topological phases in photonic crystals~\cite{Peano2016}.

This work is supported by NSFC (No.~11574100) and the National Thousand-Young-Talents Program. AH acknowledges support by DFG-SFB925 and the Hamburg centre of ultrafast imaging (CUI).

\bibliographystyle{apsrev4-1}

%

\clearpage
\onecolumngrid

\centerline{\bf Supplementary Material}
\vspace{0.1in}

\renewcommand{\thesection}{S-\arabic{section}}
\setcounter{section}{0}  
\renewcommand{\theequation}{S\arabic{equation}}
\setcounter{equation}{0}  
\renewcommand{\thefigure}{S\arabic{figure}}
\setcounter{figure}{0}  

In this supplementary material, we provide additional details on (A) interaction Hamiltonians, (B) tight-binding models, (C) effective two-band models, and (D) topological excitations.

\section*{A. Interaction Hamiltonians}
In this section, we obtain the interaction Hamiltonian for an effective two-band model and discuss its influence on the ground state. Taking the two-mode approximation, the bosonic field operator can be expanded as 
\begin{equation}
\hat{\psi}(\mathbf{r})=\sum_{i}w_{+}(\mathbf{r}-\mathbf{r}_i)\hat{b}_{+,i}+w_{-}(\mathbf{r}-\mathbf{r}_i)\hat{b}_{-,i},
\end{equation}
where $w_{\pm}(\mathbf{r}-\mathbf{r}_i)$ and $\hat{b}_{\pm,i}$ are respectively Wannier functions cetered at $\mathbf{r}_i$ and boson annihilation operators on site $i$ . Two Wannier functions are time-reversal pairs and satisfiy the relation $w_{+}(\mathbf{r})=w_{-}^*(\mathbf{r})$. Considering only the s-wave scattering among bosons, the effective interaction Hamiltonian can be written as
\begin{eqnarray}
\hat{H}_\text{int}&=&\frac{U_0}{2}\int d\mathbf{r}\,\hat{\psi}^\dag(\mathbf{r})\hat{\psi}^\dag(\mathbf{r})\hat{\psi}(\mathbf{r})\hat{\psi}(\mathbf{r})\nonumber\\
&\simeq&\frac{U_0}{2}\sum\limits_{i} \int d\mathbf{r}\, \left[w_{+}^*(\mathbf{r}-\mathbf{r}_i)\hat{b}^{\dagger}_{+,i}+w_{-}^*(\mathbf{r}-\mathbf{r}_i)\hat{b}^{\dagger}_{-,i}\right]\left[w^*_{+}(\mathbf{r}-\mathbf{r}_i)\hat{b}^{\dagger}_{+,i}+w_{-}^*(\mathbf{r}-\mathbf{r}_i)\hat{b}^{\dagger}_{-,i}\right]
\nonumber\\
&&\qquad\qquad\times \left[w_{+}(\mathbf{r}-\mathbf{r}_i)\hat{b}_{+,i}+w_{-}(\mathbf{r}-\mathbf{r}_i)\hat{b}_{-,i}\right]\left[w_{+}(\mathbf{r}-\mathbf{r}_i)\hat{b}_{+,i}+w_{-}(\mathbf{r}-\mathbf{r}_i)\hat{b}_{-,i}\right].
\label{int}
\end{eqnarray}
After expansion, we obtain 16 terms. The coefficients of part of them are zero due to symmetry requirement, momentum conservation, or orbital angular momentum conservation. Here, we consider a ferromagnetic orbital interacion which can be naturally realized experimentally when only merely $s$-wave interaction is involved, as we discussed in the main text and the following sections. The corresponding interaction Hamiltonian is thus given by 
\begin{eqnarray}
\hat{H}_\text{int}=
\frac{U_{++}}{2}\sum_i \left(\hat{b}_{+,i}^\dag \hat{b}_{+,i}^\dag \hat{b}_{+,i} \hat{b}_{+,i} + \hat{b}_{-,i}^\dag \hat{b}_{-,i}^\dag \hat{b}_{-,i} \hat{b}_{-,i}
+4 \hat{b}_{+,i}^\dag \hat{b}_{-,i}^\dag \hat{b}_{-,i} \hat{b}_{+,i}\right),
\label{int2}
\end{eqnarray}
where $U_{++}=U_0\int d\mathbf{r}|w_{+}(\mathbf{r})|^4$. Transforming the lattice model into a continuous model by $\sum_i\to\int \frac{dxdy}{\Omega}$ and $\hat{b}_{+(-),i}\to\sqrt{\Omega}\phi_{+(-)}(\mathbf{r})$, where $\Omega$ is the area of one unit cell, we get the interaction Hamiltonian in the main text
\begin{equation}
\begin{aligned}
\hat{H}_\text{int}=\frac{g}{2}\int d\mathbf{r}[\phi_+^{\dagger}(\mathbf{r})\phi_+^{\dagger}(\mathbf{r}) \phi_+(\mathbf{r})\phi_+(\mathbf{r})+\phi_-^{\dagger}(\mathbf{r})\phi_-^{\dagger}(\mathbf{r})\phi_-(\mathbf{r})\phi_-(\mathbf{r})+4\phi_+^{\dagger}(\mathbf{r})\phi_-^{\dagger}(\mathbf{r}) \phi_-(\mathbf{r})\phi_+(\mathbf{r})],
\end{aligned}
\end{equation}
where $g=\Omega U_{++}$. In the following, we consider the repulsive $s$-wave contact interaction among bosons with $g>0$.

Taking into account the single-particle Hamiltonian, we solve the ground state under a mean-field approximation. As long as the single-particle band minima are located at $\Gamma$ point. We only need to consider these two single-particle states when minimizing the total energy functional. Within the mean-field framework, we have $\langle \phi_{\pm,\mathbf{k}=\mathbf{0}}\rangle=\sqrt{N_{\pm}}e^{i\theta_{\pm}}$, where $N_{\pm}$ are the particle numbers for two different orbitals. The mean-field ground state energy is given by 
\begin{equation}
E_{\rm MF}=\frac{g}{2V}(N_+^2+N_-^2+4N_+N_-).
\end{equation}
where $V$ is the volume of the 2D system. 
The repulsive interaction among bosons favors a spontaneous symmetry broken ground state with ($N_+=N$, $N_-=0$) or ($N_+=0$, $N_-=N$).

\section*{B. Tight-binding models}

In this section, we obtain parameters for the tight-binding models via directly calculating the Wannier functions, as discussed in Ref~\cite{Uehlinger2013,Kivelson1982}. We first define two band-projected position operators for the two-dimensional system
\begin{equation}
\begin{split}
&\hat{R}_1=\sum_{\mu,\mu',\mathbf{k},\mathbf{k}^\prime}|\psi_{\mu,\mathbf{k}}\rangle \langle \psi_{\mu,\mathbf{k}}|\mathbf{b}_1\cdot\mathbf{\hat{r}}|\psi_{\mu^\prime,\mathbf{k}^\prime}\rangle \langle \psi_{\mu^\prime,\mathbf{k}^\prime}|\\
&\hat{R}_2=\sum_{\mu,\mu',\mathbf{k},\mathbf{k}^\prime}|\psi_{\mu,\mathbf{k}}\rangle \langle \psi_{\mu,\mathbf{k}}|\mathbf{b}_2\cdot\mathbf{\hat{r}}|\psi_{\mu^\prime,\mathbf{k}^\prime}\rangle \langle \psi_{\mu^\prime,\mathbf{k}^\prime}|,
\end{split}
\end{equation}
where $|\psi_{\mu,\mathbf{k}}\rangle$ represents the Bloch state in the $\mu$-th band at momentum $\mathbf{k}$, $\mathbf{b}_1$ and $\mathbf{b}_2$ are the reciprocal lattice vectors and $\mathbf{\hat{r}}=(\hat{x},\hat{y})$. Wannier functions are defined as the eigenstates of the band-projected position operators and the corresponding eigenvalues indicate the positions of Wannier functions in real space. Therefore, we obtain the Wannier functions by diagonalizing $\hat{R}_1$ and $\hat{R}_2$.

The matrix elements of the band-projected position operators can be constructed by the Bloch waves, which can be obtained by the plane wave expansion method as
\begin{equation}
\psi_{\mu,\mathbf{k}}(\mathbf{r})=\frac{1}{\sqrt{N_u\Omega}}e^{i\mathbf{k}\cdot\mathbf{r}}\sum_{n_1,n_2}c_{\mu,\mathbf{k}}^{n_1,n_2}e^{i(n_1\mathbf{b}_1+n_2\mathbf{b}_2)\cdot\mathbf{r}}.
\end{equation}
Here, the system covers $N_u$ unit cells. We thus obtain the matrix elements of the position operators by the real-space integration as
\begin{equation}
\begin{split}
&R^{(1)}=\langle\psi_{\mu,\mathbf{k}}|\mathbf{b}_1\cdot\mathbf{\hat{r}}|\psi_{\mu^\prime,\mathbf{k}^\prime}\rangle=\int_{N_u\Omega}d^2r\psi_{\mu,\mathbf{k}}^{*}(\mathbf{r})\psi_{\mu^\prime,\mathbf{k}^\prime}(\mathbf{r})\mathbf{b}_1\cdot\mathbf{r}\\
&R^{(2)}=\langle\psi_{\mu,\mathbf{k}}|\mathbf{b}_2\cdot\mathbf{\hat{r}}|\psi_{\mu^\prime,\mathbf{k}^\prime}\rangle=\int_{N_u\Omega}d^2r\psi_{\mu,\mathbf{k}}^{*}(\mathbf{r})\psi_{\mu^\prime,\mathbf{k}^\prime}(\mathbf{r})\mathbf{b}_2\cdot\mathbf{r}.
\end{split}
\end{equation}
Here, the integral is among the entire lattice with volume $V=N_u\Omega$. Suppose that there are $L=\sqrt{N_u}$ unit cells along the direction of each lattice vector, we can parametrize the momentum $\mathbf{k}$ as
\begin{equation}
\mathbf{k}=\frac{m_1}{L}\mathbf{b}_1+\frac{m_2}{L}\mathbf{b}_2,
\end{equation}
where, $m_1$ and $m_2$ should be selected as a set of integers which make the momentum $\mathbf{k}$ covers the entire first Brillouin zone. Finally, we obtain the matrix elements of $\hat{R}_1$ and $\hat{R}_2$:
\begin{equation}
\begin{split}
R^{(1)}=&(-1)^{m_1-m_1^\prime}(-1)^{m_2-m_2^\prime}\sum_{n_1,n_2,n_1^\prime,n_2^\prime}c_{\mu,\mathbf{k}}^{n_1,n_2*}c_{\mu^\prime,\mathbf{k}^\prime}^{n_1^\prime,n_2^\prime} (-1)^{L(n_1-n_1^\prime)}(-1)^{L(n_2-n_2^\prime)}\\
&\times\frac{i}{\frac{m_1-m_1^\prime}{L}+n_1-n_1^\prime}(1-\delta_{m_1,m_1^\prime}\delta_{n_1,n_1^\prime}) \delta_{m_2,m_2^\prime}\delta_{n_2,n_2^\prime}
\end{split}
\end{equation}
\begin{equation}
\begin{split}
R^{(2)}=&(-1)^{m_1-m_1^\prime}(-1)^{m_2-m_2^\prime}\sum_{n_1,n_2,n_1^\prime,n_2^\prime}c_{\mu,\mathbf{k}}^{n_1,n_2*}c_{\mu^\prime,\mathbf{k}^\prime}^{n_1^\prime,n_2^\prime} (-1)^{L(n_1-n_1^\prime)}(-1)^{L(n_2-n_2^\prime)}\\
&\times\delta_{m_1,m_1^\prime}\delta_{n_1,n_1^\prime}\frac{i}{\frac{m_2-m_2^\prime}{L}+n_2-n_2^\prime}(1-\delta_{m_2,m_2^\prime}\delta_{n_2,n_2^\prime}).
\end{split}
\end{equation}
Diagonalizing the band-projected position operators, we obtain Wannier functions and determine all tight-binding parameters.

\begin{figure}[tbp]
	\centering
	\includegraphics[width=16cm]{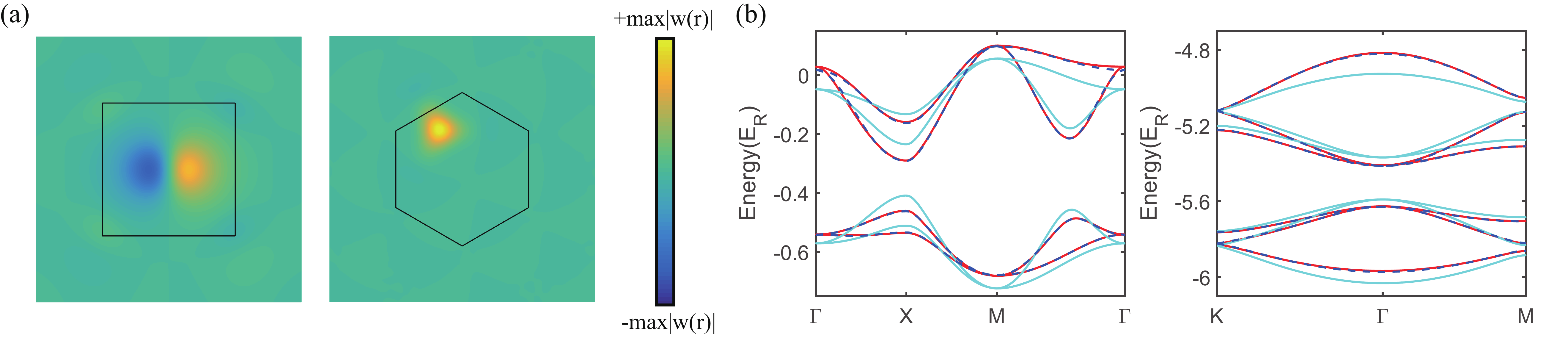}
	\caption{ (a) Selected Wannier functions for the square lattice and the honeycomb lattice, respectively. Black solid square or honeycomb denote one unit cell  of the optical lattice. (b) Single-particle energy bands. Red solid lines denote numerical results calculated from plane-wave expansion. Cyan solid lines denote energy bands derived from tight-binding models discussed in the main text and 
blue dashed lines denote results obtained from tight-binding models with more hopping terms. All hopping parameters are calculated based on the obtained Wannier functions. 
The parameter we used are (left) $V_1=1.4E_R$ and $V_2=0.1V_1$, $\delta=0.1382E_R$, $J_1=0.0555E_R$, $J_2=0.0912E_R$, $J_3=0.0158E_R$ and $J_4=-0.0098E_R$; (right) $V_1=7E_R$ and $V_2=0.1V_1$, $J=-0.1474E_R$ and $J'=-0.2584E_R$.}
\end{figure}

\section*{C. Effective two-band models}

In this section, we derive an effective two-band model for describing low-energy physics close to the QBCP. 

\subsection*{Optical lattice with $D_4$ point-group symmetry}
For the square lattice, the single-particle Hamiltonian in momentum space can be written as $H_0=\sum_\mathbf{k}\Psi_\mathbf{k}^\dag \mathcal{H}_0(\mathbf{k})\Psi_\mathbf{k}$, where $\Psi_\mathbf{k}=(\hat{p}_{Ax,\mathbf{k}},\ \hat{p}_{Ay,\mathbf{k}},\ \hat{p}_{Bx,\mathbf{k}},\ \hat{p}_{By,\mathbf{k}})^\mathrm{T}$ and 
\begin{small}
\begin{equation}
\hspace{-15mm}
\mathcal{H}_0(\mathbf{k})\!=\!
\left( \begin{smallmatrix}
\delta +2J_3\cos \left( ak_x \right) +2J_4\cos \left( ak_y \right)&		0&		2J_1\left( \cos \left( \frac{a}{2}(k_x+k_y) \right) +\cos \left( \frac{a}{2}(k_x-k_y) \right) \right)&		2J_2\left( \cos \left( \frac{a}{2}(k_x+k_y) \right) -\cos \left( \frac{a}{2}(k_x-k_y) \right) \right)\\
0&		\delta +2J_4\cos \left( ak_x \right) +2J_3\cos \left( ak_y \right)&		2J_2\left( \cos \left( \frac{a}{2}(k_x+k_y) \right) -\cos \left( \frac{a}{2}(k_x-k_y) \right) \right)&		2J_1\left( \cos \left( \frac{a}{2}(k_x+k_y) \right) +\cos \left( \frac{a}{2}(k_x-k_y) \right) \right)\\
2J_1\left( \cos \left( \frac{a}{2}(k_x+k_y) \right) +\cos \left( \frac{a}{2}(k_x-k_y) \right) \right)&		2J_2\left( \cos \left( \frac{a}{2}(k_x+k_y) \right) -\cos \left( \frac{a}{2}(k_x-k_y) \right) \right)&		-\delta+2J_3\cos \left( ak_x \right) +2J_4\cos \left( ak_y \right)&		0\\
2J_2\left( \cos \left( \frac{a}{2}(k_x+k_y) \right) -\cos \left( \frac{a}{2}(k_x-k_y) \right) \right)&		2J_1\left( \cos \left( \frac{a}{2}(k_x+k_y) \right) +\cos \left( \frac{a}{2}(k_x-k_y) \right) \right)&		0&		-\delta+2J_4\cos \left( ak_x \right) +2J_3\cos \left( ak_y \right)\\
\end{smallmatrix} \right).
\end{equation}
\end{small} 

Performing a unitary transformation $\mathcal{H}'(\mathbf{k})=\mathcal{U}^\dag\mathcal{H}(\mathbf{k})\mathcal{U}$, where the matrix $\mathcal{U}$ is given by
\begin{equation}
\mathcal{U}=\frac{1}{\sqrt{2}}\left(
\begin{matrix}
-i\sin\chi & i\sin\chi & i\cos\chi & -i\cos\chi \\
-\sin\chi & -\sin\chi & \cos\chi & \cos\chi \\
\cos\chi & \cos\chi & \sin\chi & \sin\chi \\
i\cos\chi & -i\cos\chi & i\sin\chi & -i\sin\chi
\end{matrix}\right)
\end{equation}
with $\chi=\arctan\frac{\delta+\sqrt{\delta^2+16J_2^2}}{4J_2}$, 
we transform the orbital bases into the angular momentum bases. Here, $\mathcal{U}$ contains four eigenstates at $K_1$. The last two columns of $\mathcal{U}$ are two eigenstates with lower energies. Two energy bands close to the degeneracy point are spanned by two bases: 
\begin{eqnarray}
\hat{P}_{+,\mathbf{k}_{K_1}}^{\dag}=\frac{1}{\sqrt{2}}\left( i \hat{p}_{Ax,\mathbf{k}}^{\dag}\cos \chi+ \hat{p}_{Ay,\mathbf{k}}^{\dag}\cos \chi+ \hat{p}_{Bx,\mathbf{k}}^{\dag}\sin \chi+i \hat{p}_{By,\mathbf{k}}^{\dag}\sin \chi \right),\\
\hat{P}_{-,\mathbf{k}_{K_1}}^{\dag}=\frac{1}{\sqrt{2}}\left( -i \hat{p}_{Ax,\mathbf{k}}^{\dag}\cos \chi+ \hat{p}_{Ay,\mathbf{k}}^{\dag}\cos \chi+ \hat{p}_{Bx,\mathbf{k}}^{\dag}\sin \chi-i \hat{p}_{By,\mathbf{k}}^{\dag}\sin \chi \right),
\end{eqnarray} 
where $\mathbf{k}_{K_1}=\mathbf{k}-(\pi/a,\pi/a)$. We thus obtain an effective two-band model given by $\tilde{H}_0=\sum_{\mathbf{k}}{\left( \hat{P}_{+,\mathbf{k}}^{\dag},\hat{P}_{-,\mathbf{k}}^{\dag} \right)}\tilde{\mathcal{H}}_0\left( \mathbf{k} \right) \left(\hat{P}_{+,\mathbf{k}}, \hat{P}_{-,\mathbf{k}} \right)^T$. Expanding the matrix $\tilde{\mathcal{H}}_0(\mathbf{k})$ upto the second order of $\mathbf{k}$, we obtain
\begin{equation}
\label{eff_QBCP}
\tilde{\mathcal{H}}_0\left( \mathbf{k} \right) =
\left( \begin{matrix}
t_0(k_x^{2}+k_y^{2})&		t_1(k_x^2-k_y^2)-it_2k_xk_y\\
t_1(k_x^2-k_y^2)+it_2k_xk_y&		t_0(k_x^{2}+k_y^{2})\\
\end{matrix} \right),
\end{equation}
where $t_0=\frac{a^2(J_3+J_4+J_2\sin2\chi)}{2}$, $t_1=\frac{a^2(J_4-J_3)\cos2\chi}{2}$, and $t_2=a^2J_1\sin2\chi$.

For the square optical lattices we considered in the main text, using the matrix $\mathcal{U}$ we derive two Wannier functions related to two bases $\hat{P}_{\pm,\mathbf{k}}$ as
\begin{eqnarray}
w_{+}(\mathbf{r}) =\frac{1}{\sqrt{2}}\{[i w_{p_{Ax}}( \mathbf{r} ) + w_{p_{Ay}}( \mathbf{r})]\cos \chi + [w_{p_{Bx}}( \mathbf{r}) +i w_{p_{By}}( \mathbf{r})]\sin \chi\}, \\
w_{-}(\mathbf{r}) =\frac{1}{\sqrt{2}}\{[-i w_{p_{Ax}}( \mathbf{r} ) +w_{p_{Ay}}( \mathbf{r})]\cos\chi + [w_{p_{Bx}}( \mathbf{r}) -i w_{p_{By}}( \mathbf{r})]\sin\chi\},
\end{eqnarray}
where $w_{p_{O\mu}}$ ($O=A,B$, $\mu=x,y$) is the Wannier function of the $p_{\mu}$ orbital at $O$ site. We define $U_{A(B)1}=U_0\int d\mathbf{r}|w_{p_{A(B)x}}(\mathbf{r})|^4=U_0\int d\mathbf{r}|w_{p_{A(B)y}}(\mathbf{r})|^4$ and $U_{A(B)2}=U_0\int d\mathbf{r}|w_{p_{A(B)x}}(\mathbf{r})|^2|w_{p_{A(B)y}}(\mathbf{r})|^2$. Substituting two Wannier functions into the Hamiltonian of Eq.~(\ref{int}) and applying the relation $U_{A(B)1}=3U_{A(B)2}$ as obtained in Ref.~\cite{Liu2006}, we confirmed that the interaction Hamiltonian is same as that shown in Eq.~(\ref{int2}) where $\hat{b}_{\pm}$ are replaced by $\hat{P}_{\pm}$. The interacting coefficient is obtained as
$U_{++}=[(U_{A1}+U_{A2})\cos^4\chi+(U_{B1}+U_{B2})\sin^4\chi]/2=\frac{2}{3}(\cos^4\chi U_{A1}+\sin^4\chi U_{B1})$.

\subsection*{Optical lattice with $D_6$ point-group symmetriy}
For the lattice with $D_6$ point-group symmetry discussed in the main text, we propose a tight-binding model with six $s$ orbitals to describe the system. The single-particle Hamiltonian in momentum space can be written as $H_0=\sum_\mathbf{k}\Psi_\mathbf{k}^\dag \mathcal{H}_0(\mathbf{k})\Psi_\mathbf{k}$, where $\Psi_\mathbf{k}=(\hat{s}_{1,\mathbf{k}}\ \hat{s}_{2,\mathbf{k}}\ \hat{s}_{3,\mathbf{k}}\ \hat{s}_{4,\mathbf{k}}\ \hat{s}_{5,\mathbf{k}}\ \hat{s}_{6,\mathbf{k}})^\mathrm{T}$ and 
\begin{equation}
\begin{aligned}
\mathcal{H}_0(\mathbf{k})=J
\left( \begin{matrix}
0&e^{i\mathbf{e}_3\cdot\mathbf{k}}&0&\gamma e^{i\mathbf{e}_2\cdot\mathbf{k}}&0&e^{i\mathbf{e}_1\cdot\mathbf{k}}\\
e^{-i\mathbf{e}_3\cdot\mathbf{k}}&0&e^{-i\mathbf{e}_2\cdot\mathbf{k}}&0&\gamma e^{-i\mathbf{e}_1\cdot\mathbf{k}}&0\\
0&e^{i\mathbf{e}_2\cdot\mathbf{k}}&0&e^{i\mathbf{e}_1\cdot\mathbf{k}}&0&\gamma e^{i\mathbf{e}_3\cdot\mathbf{k}}\\
\gamma e^{-i\mathbf{e}_2\cdot\mathbf{k}}&0&e^{-i\mathbf{e}_1\cdot\mathbf{k}}&0&e^{-i\mathbf{e}_3\cdot\mathbf{k}}&0\\
0&\gamma e^{i\mathbf{e}_1\cdot\mathbf{k}}&0&e^{i\mathbf{e}_3\cdot\mathbf{k}}&0&e^{i\mathbf{e}_2\cdot\mathbf{k}}\\
e^{-i\mathbf{e}_1\cdot\mathbf{k}}&0&\gamma e^{-i\mathbf{e}_3\cdot\mathbf{k}}&0&e^{-i\mathbf{e}_2\cdot\mathbf{k}}&0\\
\end{matrix} \right),
\end{aligned}
\end{equation}
where $\mathbf{k}=(k_x,k_y)$, $\mathbf{e}_1=a_0(1,0)$, $\mathbf{e}_2=a_0(-\frac{1}{2},\frac{\sqrt{3}}{2})$, and $\mathbf{e}_3=a_0(-\frac{1}{2},-\frac{\sqrt{3}}{2})$. $a_0=\frac{4\pi}{3\sqrt{3}k_L}$ is the distance between two nearst-neighbor lattice sites.

\begin{figure}[b]
	\centering
	\includegraphics[width=3cm]{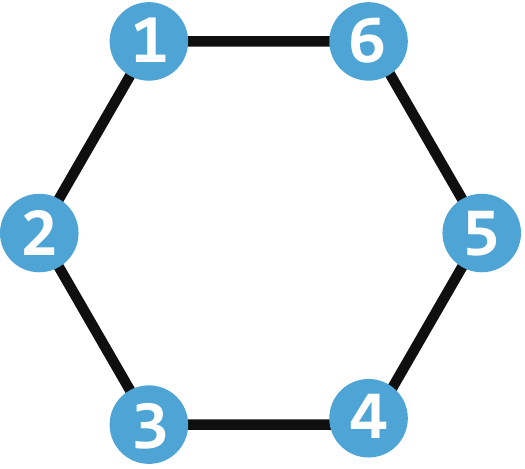}
	\caption{Schematic picture of the optical lattice with $D_6$ point-group symmetry. The numbers label different sites in one unit cell.}
\end{figure}

To obtain the low-energy effective Hamiltonian close to the QBCP, we first change the bases from six spatially separated $s$-orbitals to a set of orbitals named $s$, $p_x$, $p_y$, $d_{x^2-y^2}$, $d_{xy}$ and $f$ orbitals. They can be constructed by applying the projection operators to calculate the irreducible representation bases. We thus obtain them as superpositions of six $s$ orbitals as
\begin{equation}
\begin{aligned}
\mathbf{u}_s&=\frac{1}{\sqrt{6}}
\left[\begin{matrix}
1&1&1&1&1&1
\end{matrix}\right]^\mathrm{T},\\
\mathbf{u}_{p_x}&=\frac{1}{2\sqrt{3}}
\left[\begin{matrix}
-1&-2&-1&1&2&1
\end{matrix}\right]^\mathrm{T},\\
\mathbf{u}_{p_y}&=\frac{1}{2}
\left[\begin{matrix}
1&0&-1&-1&0&1
\end{matrix}\right]^\mathrm{T},\\
\mathbf{u}_{d_{x^2-y^2}}&=\frac{1}{2\sqrt{3}}
\left[\begin{matrix}
-1&2&-1&-1&2&-1
\end{matrix}\right]^\mathrm{T},\\
\mathbf{u}_{d_{xy}}&=\frac{1}{2}
\left[\begin{matrix}
-1&0&1&-1&0&1
\end{matrix}\right]^\mathrm{T},\\
\mathbf{u}_f&=\frac{1}{\sqrt{6}}
\left[\begin{matrix}
1&-1&1&-1&1&-1
\end{matrix}\right]^\mathrm{T}.\\
\end{aligned}
\end{equation}

For the $\Gamma$ point, the little group coincides with the point group $D_6$. Therefore, $\mathbf{u}_{o}$ with $o$=$s$, $p_x$, $p_y$, $d_{x^2-y^2}$, $d_{xy}$ and $f$, are eigenstates of $\mathcal{H}_0(\mathbf{k})$ at $\Gamma$. The corresponding eigenvalues are $\varepsilon_s=(2+\gamma)J$ for $s$ orbital, $\varepsilon_f=-(2+\gamma)J$ for $f$ orbital, $\varepsilon_d=(\gamma-1)J$ for $d$ orbitals and $\varepsilon_p=-(\gamma-1)J$ for $p$ orbitals.
Close to $\Gamma$ point, the middle four bands mainly spanned by $p_x$, $p_y$, $d_{x^2-y^2}$ and $d_{xy}$ orbitals. The corresponding Hamiltonian becomes $H_0=\sum_\mathbf{k}\Psi_\mathbf{k}^\dag \mathcal{H}_0(\mathbf{k})\Psi_\mathbf{k}$, where $\Psi_\mathbf{k}=(\hat{p}_{+,\mathbf{k}}\ \hat{d}_{+,\mathbf{k}}\ \hat{p}_{-,\mathbf{k}}\ \hat{d}_{-,\mathbf{k}})^\mathrm{T}$, $\hat{p}_{+,\mathbf{k}}^\dag=\frac{1}{\sqrt{2}}( \hat{p}_{x,\mathbf{k}}^\dag+i\hat{p}_{y,\mathbf{k}}^\dag)$, $\hat{p}_{-,\mathbf{k}}^\dag=\frac{1}{\sqrt{2}}( \hat{p}_{x,\mathbf{k}}^\dag-i\hat{p}_{y,\mathbf{k}}^\dag)$, $\hat{d}_{+,\mathbf{k}}^\dag=\frac{1}{\sqrt{2}}( \hat{d}_{x^2-y^2,\mathbf{k}}^\dag+i\hat{d}_{xy,\mathbf{k}}^\dag)$ and $\hat{d}_{-,\mathbf{k}}^\dag=\frac{1}{\sqrt{2}}( \hat{d}_{x^2-y^2,\mathbf{k}}^\dag-i\hat{d}_{xy,\mathbf{k}}^\dag)$. Keeping upto the second order of momentum, $\mathcal{H}_0(\mathbf{k})$ is written as
\begin{equation}
\mathcal{H}_0(\mathbf{k})=
\left( \begin{matrix}
D-B\left( k_{x}^{2}+k_{y}^{2} \right)&		-iA\left( k_x+ik_y \right)&		-C\left( k_x-ik_y \right) ^2&		0\\
iA\left( k_x-ik_y \right)&		-D+B\left( k_{x}^{2}+k_{y}^{2} \right)&		0&		C\left( k_x+ik_y \right) ^2\\
-C\left( k_x+ik_y \right) ^2&		0&		D-B\left( k_{x}^{2}+k_{y}^{2} \right)&		-iA\left( k_x-ik_y \right)\\
0&		C\left( k_x-ik_y \right) ^2&		iA\left( k_x+ik_y \right)&		-D+B\left( k_{x}^{2}+k_{y}^{2} \right)\\
\end{matrix} \right),
\end{equation}
where $A=-\frac{a_0J(\gamma+2)}{2}$, $B=-\frac{a_0^2J(\gamma-1)}{4}$, $C=-\frac{a_0^2J(\gamma+2)}{8}$ and $D=-J(\gamma-1)$.

When $\gamma>1$, the upper two bands are spanned by two $p$-orbitals. Projecting out two $d$-orbitals via a second order perturbation, we obtain a two-band model as $\tilde{H}_0=\sum_{\mathbf{k}}(\hat{p}_{+,\mathbf{k}},\hat{p}_{-,\mathbf{k}}^{\dag}) \tilde{\mathcal{H}}_0(\mathbf{k})(\hat{p}_{+,\mathbf{k}},\hat{p}_{-,\mathbf{k}})^\mathrm{T}$, where
\begin{eqnarray}
\tilde{\mathcal{H}}_0(\mathbf{k})&=&
\left( \begin{matrix}
[\mathcal{H}_0]_{33}&		[\mathcal{H}_0]_{34}\\
[\mathcal{H}_0]_{43}&		[\mathcal{H}_0]_{44}\\
\end{matrix} \right)
+\frac{1}{\varepsilon_p-\varepsilon_d}
\left(
\begin{matrix}
[\mathcal{H}_0]_{31}[\mathcal{H}_0]_{13}+[\mathcal{H}_0]_{32}[\mathcal{H}_0]_{23}& [\mathcal{H}_0]_{31}[\mathcal{H}_0]_{14}+[\mathcal{H}_0]_{32}[\mathcal{H}_0]_{24}\\
[\mathcal{H}_0]_{41}[\mathcal{H}_0]_{13}+[\mathcal{H}_0]_{42}[\mathcal{H}_0]_{23}& [\mathcal{H}_0]_{41}[\mathcal{H}_0]_{14}+[\mathcal{H}_0]_{42}[\mathcal{H}_0]_{24}
\end{matrix}\right)\nonumber\\
&=&
\left( \begin{matrix}
t_0(k_x^{2}+k_y^{2})&		t_1(k_x^2-k_y^2)-it_2k_xk_y\\
t_1(k_x^2-k_y^2)+it_2k_xk_y&		t_0(k_x^{2}+k_y^{2})\\
\end{matrix} \right),
\end{eqnarray}
where $t_0=\frac{a_0^2J(\gamma^2-8\gamma-2)}{8(\gamma-1)}$ and $t_1=t_2=\frac{a_0^2J(\gamma+2)}{8}$.

We further obtain two Wannier functions related to $\hat{p}_{\pm,\mathbf{k}}$ as
\begin{eqnarray}
w_{+}(\mathbf{r}) =\frac{1}{\sqrt{6}}[e^{i\frac{2\pi}{3}}w_{s_1}(\mathbf{r})-w_{s_2}(\mathbf{r})+e^{-i\frac{2\pi}{3}}w_{s_3}(\mathbf{r})+e^{-i\frac{\pi}{3}}w_{s_4}(\mathbf{r})+w_{s_5}(\mathbf{r})+e^{i\frac{\pi}{3}}w_{s_6}(\mathbf{r})],\\
w_{-}(\mathbf{r}) =\frac{1}{\sqrt{6}}[e^{-i\frac{2\pi}{3}}w_{s_1}(\mathbf{r})-w_{s_2}(\mathbf{r})+e^{i\frac{2\pi}{3}}w_{s_3}(\mathbf{r})+e^{i\frac{\pi}{3}}w_{s_4}(\mathbf{r})+w_{s_5}(\mathbf{r})+e^{-i\frac{\pi}{3}}w_{s_6}(\mathbf{r})].
\end{eqnarray}
 Substituting two Wannier functions into the Hamiltonian of Eq.~(\ref{int}), we confirmed that the interaction Hamiltonian is same as that shown in Eq.~(\ref{int2}) where $\hat{b}_{\pm}$ are replaced by $\hat{p}_{\pm}$. The interacting coefficient is obtained as $U_{++}=U_s/6$, 
where $U_{s}=U_0\int d\mathbf{r}|w_{s_l}(\mathbf{r})|^4$, $l$ labels different $s$ orbitals in the unit cell with integer $l=1,2,. ..,6$. 

\section*{D. Topological excitations}

In this section, we calculate the bosonic excitations on top of the time-reversal symmetry broken condensate. Within the Bogoliubov approximation, the Hamiltonian can be rewritten as
 $H=\frac{1}{2}\sum_{\mathbf{k}}(\delta{\Psi}_{\mathbf{k}}^{\dagger},\delta {\Psi}_{-\mathbf{k}}^\mathrm{T})\mathcal{H}_\textrm{BdG}(\mathbf{k}) (\delta\Psi_{\mathbf{k}}, \delta\Psi_{-\mathbf{k}}^{\dagger\mathrm{T}})$, where $\Psi_{\mathbf{k}}$ indicate the bases used in the tight-binding model, $\mathcal{H}_\textrm{BdG}(\mathbf{k})$ has the form of
\begin{equation}
\label{BdG}
\mathcal{H}_\textrm{BdG}(\mathbf{k})=\left(
\begin{matrix}
\mathcal{M}_\mathbf{k}&\mathcal{N}_\mathbf{k}\\
\mathcal{N}_\mathbf{-k}^*&\mathcal{M}_\mathbf{-k}^*
\end{matrix}
\right).
\end{equation}

To satisfy the bosonic commutation relation, the BdG Hamiltionian should be diagonalized by a paraunitary matrix $T_\mathbf{k}^\dag \mathcal{H}_\textrm{BdG}(\mathbf{k}) T_\mathbf{k}=E_\mathbf{k}$, where $T^{\dagger}_{\mathbf{k}}\tau_zT_{\mathbf{k}}=\tau_z$ and $\tau_z=\sigma_z\otimes\mathbbm{1}_{n\times n}$, $\mathbbm{1}_{n\times n}$ is an identity matrix and $n$ is the number of bands with $n=4$ for the square lattice and $n=6$ for the honeycomb lattice. In the numerical calculation, we diagonalize $\tau_z\mathcal{H}_\textrm{BdG}(\mathbf{k})$ for all $\mathbf{k}$ to obtain the excitation spectra of quasiparticles.

\begin{figure}[htbp]
	\centering
	\includegraphics[width=16cm]{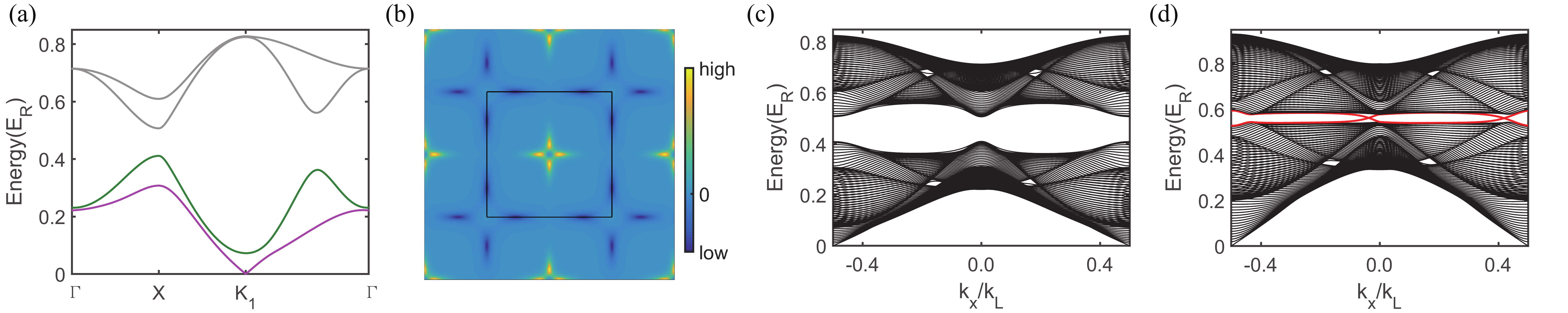}
	\caption{(a) Bosonic excitation spectra with $\delta=0.1382E_R$, $J_1=0.0555E_R$, $J_2=0.0912E_R$, $J_3=0.0158E_R$, $J_4=-0.0098E_R$ and $U_{A1}\rho=U_{B1}\rho=0.2E_R$. (b) Berry curvature for the lowest excitation band. Here, solid square denotes the first Brillouin zone. (c, d) Excitation spectra for a finite system with (c) $U_{A1}\rho=U_{B1}\rho=0.2E_R$ and (d) $U_{A1}\rho=U_{B1}\rho=0.6E_R$ . The red solid lines indicate the edge states. }
	\label{sfig3}
\end{figure}

For the square lattice, $\Psi_{\mathbf{k}}=(\hat{p}_{Ax,\mathbf{k}},\ \hat{p}_{Ay,\mathbf{k}},\ \hat{p}_{Bx,\mathbf{k}},\ \hat{p}_{By,\mathbf{k}})^\mathrm{T}$.  When bosons condensed at the QBCP, we have $\langle \Psi_{\mathbf{k}=K_1} \rangle=\frac{\sqrt{N}}{\sqrt{2}}(i\cos\chi,\ \cos\chi,\ \sin\chi,\ i\sin\chi)^\mathrm{T}$, where $\chi$ is determined by the method of simulated annealing. Although bosons condense at momentum $K_1(\frac{\pi}{a},\frac{\pi}{a})$ rather than $\Gamma(0,0)$, the BdG Hamiltionian still has the same form of Eq.~(\ref{BdG}) because of the relation $\mathcal{M}_\mathbf{-k}=\mathcal{M}_{2K_1-\mathbf{k}}$. We find that $\mathcal{M}_{\mathbf{k}}=\mathcal{H}_0(\mathbf{k})-E_{\rm QBCP} 1_{4\times 4}+\frac{\rho}{3}\text{diag}\left( u_A,u_A,u_B,u_B \right)$, where $u_A=4U_A\cos ^2\chi -2U_A\cos ^4\chi -2U_B\sin ^4\chi $, $u_B=4U_B\sin ^2\chi -2U_A\cos ^4\chi -2U_B\sin ^4\chi $ and $E_{\rm QBCP}$ is the single-particle energy at the QBCP, and
\begin{equation}
\mathcal{N}_{\mathbf{k}}=\frac{\rho}{3}\left( \begin{matrix}
-U_A\cos ^2\chi&		iU_A\cos ^2\chi&		0&		0\\
iU_A\cos ^2\chi&		U_A\cos ^2\chi&		0&		0\\
0&		0&		U_B\sin ^2\chi&		iU_B\sin ^2\chi\\
0&		0&		iU_B\sin ^2\chi&		-U_B\sin ^2\chi\\
\end{matrix} \right).
\end{equation}
Here, $\rho$ is the averaged boson number per unit cell and $U_A$ and $U_B$ is the contact interactions for $p$ orbital at site A and B.
\begin{figure}[htbp]
	\centering
	\includegraphics[width=12cm]{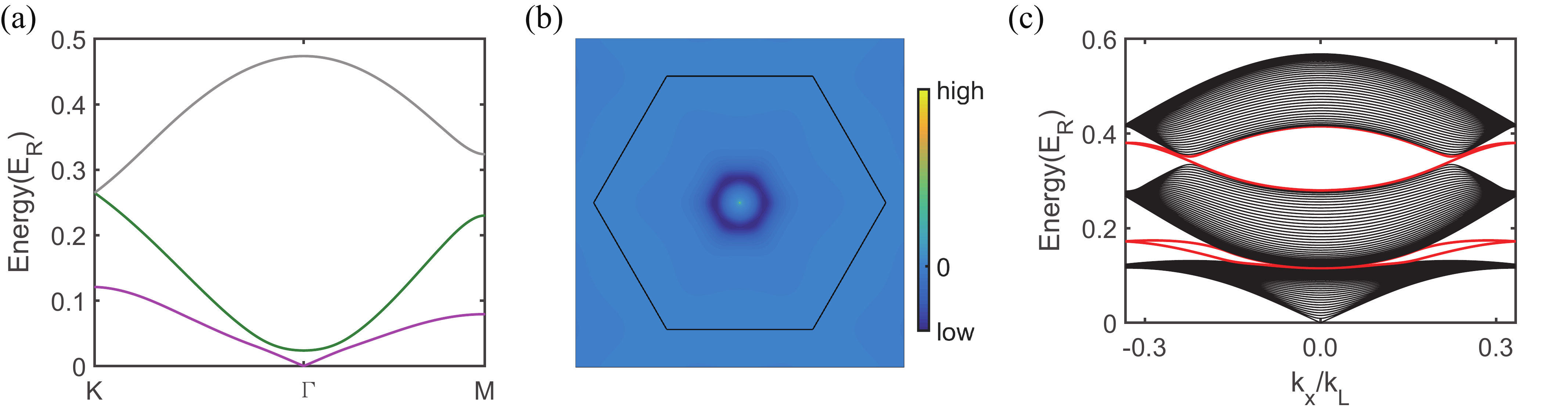}
	\caption{(a) Bosonic excitation spectra with $J=-0.15E_R$, $J'=-0.45E_R$ and $U_s\rho=0.15E_R$. (b) Berry curvature for the lowest excitation band. Here, solid honeycomb denotes the first Brillouin zone. (c) Excitation spectra for a finite system with $J=-0.15E_R$, $J'=-1.5E_R$ and $U_s\rho=0.9E_R$. Red solid lines represent edge modes.}
	\label{sfig4}
\end{figure}

For the honeycomb lattice, $\Psi_{\mathbf{k}}=(\hat{s}_{1,\mathbf{k}},\ \hat{s}_{2,\mathbf{k}},\ \hat{s}_{3,\mathbf{k}},\ \hat{s}_{4,\mathbf{k}},\ \hat{s}_{5,\mathbf{k}},\ \hat{s}_{6,\mathbf{k}})^T$. We consider to load bosons into the QBCP at minimum of the fourth and the fifth bands. Due to the ferromagnetic orbital interaction, when bosons condense the time-reversal symmetry will be broken. For the case of $\gamma>1$, we choose one degenerate state with the order parameter $\langle\Psi_{\mathbf{k}=\mathbf{0}}\rangle=\frac{\sqrt{N}}{\sqrt{6}}(e^{i2\pi/3},\ -1,\ e^{-i2\pi/3},\ e^{-i\pi/3},\ 1,\ e^{i\pi/3})^{\mathrm{T}}$. The BdG Hamiltionian is given by Eq.~(\ref{BdG}) with $\mathcal{M}_{\mathbf{k}}=\mathcal{H}_0\left( \mathbf{k} \right) -E_{\rm QBCP} \mathbbm{1}_{6\times 6}+\frac{U\rho}{6}\mathbbm{1}_{6\times 6}$ and $\mathcal{N}_{\mathbf{k}}=\frac{U\rho}{6}\text{diag}(e^{-i2\pi/3},\ 1,\ e^{i2\pi/3},\ e^{-i2\pi/3},\ 1,\ e^{i2\pi/3})$.

Diagonalize the BdG Hamiltonian, we obtain bosonic excitation spectra. To characterize the topological feature of the bosonic excitations, we numerically calculate the topological invariant. For the $j$-th band, it is defined as
\begin{equation}
\mathcal{C}_j=\frac{1}{2\pi}\int d\mathbf{k} B_j(\mathbf{k}),
\end{equation}
where the Berry curvature $B_j(\mathbf{k})$ reads
\begin{equation}
B_j(\mathbf{k})=\partial_{k_x}A_{j,y}(\mathbf{k})-\partial_{k_y}A_{j,x}(\mathbf{k}),\quad A_{j,\nu}(\mathbf{k})= i{\rm Tr}[\Gamma_j \tau_zT^{\dagger}_{\mathbf{k}}\tau_z\partial_{k_{\nu}}T_{\mathbf{k}}].
\end{equation}
Here, $\Gamma_j$ is a diagonal matrix with the $j$-th diagonal term equal to 1 and other terms are 0.

Due to the bulk-boundary correspondence, nonzero topological invariant indicates the existence of the edge modes. For the finite system, we consider a strip geometry with the periodic (open boundary condition along  $x$ ($y$) direction. Numerically, one unit cell of the finite system covers 40 unit cells of the optical lattice. Since a $2\pi$ Berry flux is encoded in the QBCP, when the double degeneracy is lifted by the interaction among bosons, nonzero Berry curvature is generated for the excitations, which is illustrated in Fig.~\ref{sfig3}(b) and Fig.~\ref{sfig4}(b). 

For the $D_4$ symmetric lattice, we find that with a weak interaction lowest two excitation bands shown in Fig.~\ref{sfig3}(a) are topological trivial, although Berry curvatures close to $K_1$  and $\Gamma$ points are nonzero. This is because the single-particle energy bands are degenerate at both $K_1$ and $\Gamma$ points. The time-reversal symmetry broken condensate lift the degeneracy at both points. The generated Berry curvatures cancel each other leading to zero topological invariants. When we further increase the $s$-wave interaction among bosons, we find that interaction-induced band inversion appears among the second and the third excitation bands, leading to topological excitations and in-gap edge modes, which is shown in Fig.~\ref{sfig3}(d).

For the $D_6$ symmetric lattice, we obtain similar results as the $D_4$ symmetric lattice when $\gamma<1$. While for the case of $\gamma>1$, there is only one degeneracy point among the fourth and the fifth single-particle energy bands. The interaction-induced time-reversal symmetry broken condensate generates a topological gap at the $\Gamma$ point. Meanwhile, a topological bulk gap is opened close to Dirac cones at $K$ points. For a finite system, we confirm that there are in-gap edge modes, which is shown in Fig.~\ref{sfig4}(c).

\end{document}